# A Retrieval Mechanism for Multi-versioned Digital Collection Using TAG

Dr. M. Thangaraj[#1], V. Gayathri[*2]

[#]Associate Professor, Department of Computer Science, Madurai Kamaraj University, Madurai, TN, India.
[*]Research Scholar, Department of Computer Science, Madurai Kamaraj University, Madurai, TN, India.

*Abstract*— **As the marvellous growth of the digital library in each year, the problems with indexing and searching a digital library is increased in a high rate. When the researchers search for the earlier versions, only a few recent versions in the back volumes can be retrieved soon. It is unpredictable that researchers require the earlier versions in a specific boundary. In order to facilitate the researchers, who may access any version at any time, we propose a VTAG technique for indexing. Our experiments indicate that the proposed retrieval technique, VTAG, effectively retrieves any version in considerable amount of time than the existing method.**

*Keywords*— **Version Retrieval, TAG tree, VTAG, Digital library, Information retrieval, Version Management.**

## I. INTRODUCTION

With the invention and widespread use of the internet, the amount of available information on the web and in digital libraries has been increasing at a high rate. Individuals have struggled to keep up with the increased information. Correspondingly, extensive methods and algorithms to reduce and/or rank this digital information are being researched to assist users in their searches [2].

When compared to other peoples, the persons who access the library very often is researchers. For the research, mostly the literatures are needed in wide range. Getting the latest version/edition is easier, than the old one. To access the versioned object, an effective indexing system is needed. The versions in the digital library database are nothing but different editions of a book, series of articles on a topic etc.

The problem of managing multiple versions of documents is present in many applications [12] and poses new research challenges. Traditional application domains that rely on version management, such as software configuration and cooperative work, increasingly use multi-versioned information as they migrate to a web-based environment.

The ideal solution is a version management system supporting multiple versions of the same document, while avoiding duplicate storage of their shared segments. To assure link permanence, professionally managed sites and content providers will have to rely on document versioning. In fact, we might soon see 'e-permanence' standards established for critical web sites of public interest [8]. In this paper, we present an index based model that controls the access of versioned objects in the digital library database.

The remainder of the paper is organized as follows: Section 2 is devoted to the issues relevant to version management. In Section 3, we describe the architecture for version control. Section 4 shows our performance evaluation result. Finally, in Section 5 we present conclusion.

## II. RELATED WORK

There are many digital collection search systems, such as Google Scholar [6], IEEE Xplore [7], and etc., available online. These systems produce results based on the relevancy to the query term and/or the importance of the papers. Even though these systems have provisions to search for the versions, it requires the exact keyword phrase.

In [3] two approaches are introduced, namely Edit - Based Version Retrieval (EBVR) and Reference - Based Version Retrieval (RBVR). In EBVR, the document objects are separated from the edit script. The edit-script is a file which maintains the logical order of the document and helps in identifying the useful pages per versions. A given version is reconstructed by first visiting the edit scripts to identify the objects valid for this version (in their appropriate document order). Then, the data pages containing the actual objects are retrieved.

In RBVR, reference records of a version always refer to its previous version, which in turn might refer to its previous version. Therefore, reference records are logical and may be indirect. It does not assume any physical representation of the versioned object.

Mostly the keyword based retrieval is focused more for the research in the field of Information Retrieval. In which they focus on the similarity/matching between the query term and the documents [4, 5, 1]. There is no much focus to address the problem of retrieving the back volumes effectively. Yet the search systems struggle for even the general retrieval [9]; search systems produce a plenty of result pages, in which most of the results are not interested to the users, and also diffuses the topic.

TAG [11] provides a way to minimise the topic diffusion and returns the relevant documents. It uses the Context-based Search and TAG-tree for effective indexing. TAG-tree is a combination of B+-tree and list. The Contexts and the publications are mapped into the tree. This process is independent of query and it is pre-executed. When a query is posted, it is searched against the TAG-tree and the synonyms of it are also searched. Thus the query is treated as a context and not as a set of keywords. Thus it avoids the topic diffusion. Also the structure helps more in indexing and retrieval. The reengineered TAG-tree architecture will be suitable for storing and retrieving versioned documents effectively.

### III. VTAG-INDEXER

This section shows how the indexing structure TAG-tree can be applied to address the issue of handling multiple editions. The TAG-tree has a list in the buckets of each leaf node, which is filled with the synonyms. The internal node buckets has information in the form of patterns, i.e., <prefix> <context> <suffix>. The <context> tuple have significant terms, where as the surrounding terms are in <prefix> and <suffix> tuples. With the help of patterns, as well as the way of indexing in the TAG-tree, searching is more focused; in turn that avoids the topic diffusion. The same TAG-tree is used here in VTAG indexing model, where the list is replaced by a hash table. The hash table is used to hold the version information. Fig. 1. shows the overview of the VTAG Indexer.

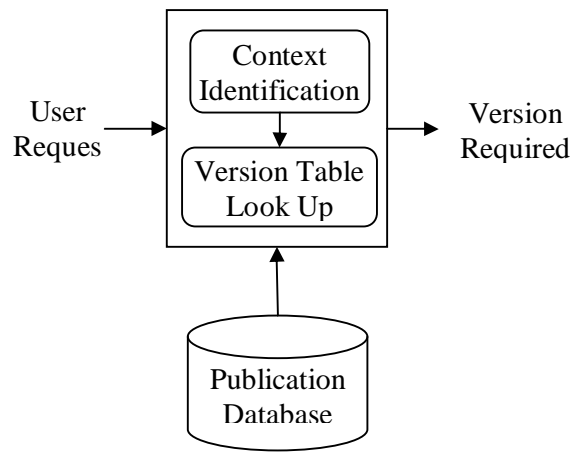

Fig. 1 Architecture of VTAG-Indexer

Each book in the publication database is pre-processed. Only the vital information of the book such as title, author, edition, publisher, year and so on., are considered for pre-processing. The processes like tokenization, stop words removal are carried over these extracted information. Then the patterns are created, based on which the books are classified. The classified books are finally indexed and mapped in VTAG-tree. The work flow of VTAG-Indexer is shown in fig. 2.

When the query is post by the user, it is searched in the nodes of the VTAG-tree for the identification of the exact book. Then the version queried is retrieved from the version table. While searching the tree, the query is treated in the form of pattern. An overview of VTAG-tree is shown in fig. 3. The books are indexed in the buckets of the B+-tree in context-based method (refer TAG for more details). The internal nodes of the $B^+$-tree is filled with only the indexing information, where as the leaf nodes have the other information like author, publications, and so on. Also the bucket of the leaf node has a pointer to a hash table, which is filled with the various versions of the book.

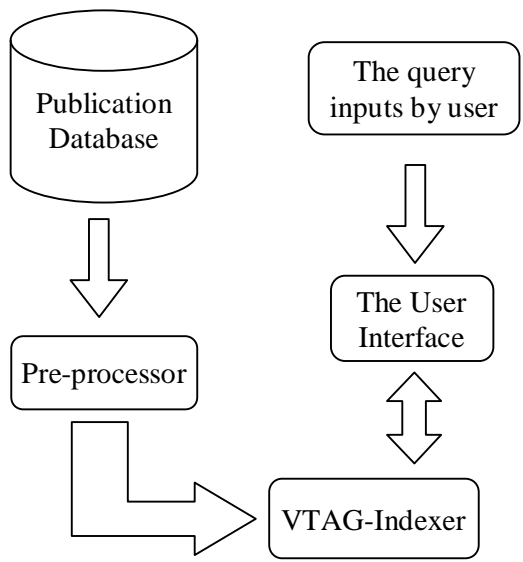

Fig. 2 The workflow of VTAG-Indexer

Any effective hash function can be used in the hash table. When a query is posed, it is searched in the VTAG-tree, finds the exact leaf node which contains the details of the book that is enquired. From the leaf, it reaches the particular hash table which has the various versions/editions of that book. The version id of the query is matched against the versions in the hash table by means of hash function. Thus it returns the exact version of the book queried. If the version queried is not available, then the latest version, i.e., the object whose version id is high, is returned to the user.

### A. Algorithm

This section shows the algorithm used for the retrieval from the multi-versioned publications. The following VersionBasedRetrieval algorithm gets the input, query as the term to be searched and the version id; produces the resultant document of the queried version;

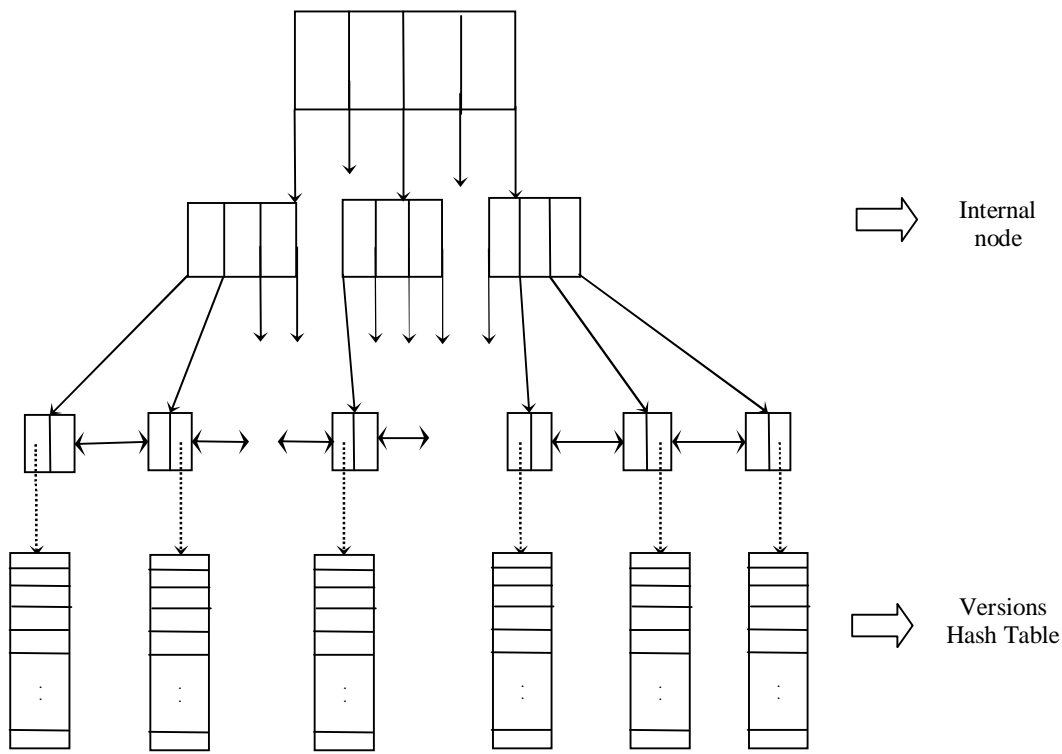

Fig. 3 Overview of VTAG-tree

**Algorithm** : **VersionBasedRetrieval**
**Data Structure:** See Fig.3.
**Input** : *qb* - Query,

*Vid* - Version Id.
**Output** : *Dm* - the resultant document of version *m*.

Let
$D_i$ - $i^{th}$ version of the document $D$,
$n$ - the latest version of document $D$,
$vl$ - version list pointer of node result,
$v$ - hash value; $1 \leq v \leq n$,
$n$ - the latest version of document $D$,
$m$ - queried version $vid$ or the latest version of the document.

findBV(Query $qb$, int $vid$)
{
    //Identify the Book through the context
    $result = find(qb, head)$;
    //locates the exact leaf of VTAG-tree.
    $vtp = result \rightarrow vl$;
    //redirects to the appropriate version table.
    $v = hash(vid)$;
    //compute the hash value of the given
    //version id.
    if($v$)
        return $D_v$;
    else
        return $D_n$;
    //if it is avail, returns the queried version.
    //else, returns the latest version.
}

The function find calls the algorithm Context Identification in TAG, which identifies and returns the queried node. Refer the module TAG Retriever, for more information on how the searching carried over the tree nodes. hash refers the hash function used in the version table.

## IV. PERFORMANCE ANALYSIS

In this section we present the results of experiments to show the efficiency of VTAG-Indexer. Various experiments were conducted in this work, but only vital information is presented here. To study the performance of this model, we have taken 200 documents. Each document has 100 versions, with 4MB. Each version changes minimum 20% from the previous version. To analyse the performance of the VTAG, it is compared with the EBVR and RBVR.

All these models were implemented using Java. The experiments were conducted on Processor Intel Core i7 @ 2.30GHz with 8 GB RAM and 1 TB hard disk, running Windows 7.

### A. Object Retrieval

When the size of the collection increases, it becomes a hectic task of finding the particular document (object). Fig. 4. shows the time taken retrieve the document. Since our VTAG model uses the context - based approach, as well the hierarchical approach, it takes, a minimum time to locate the document. Both EBVR and RBVR searches linearly in the collection, and uses the traditional keyword-based search mechanism. Thus the number of comparisons to locate the specified document is quite large in the existing system, which in turn increases the retrieval time.

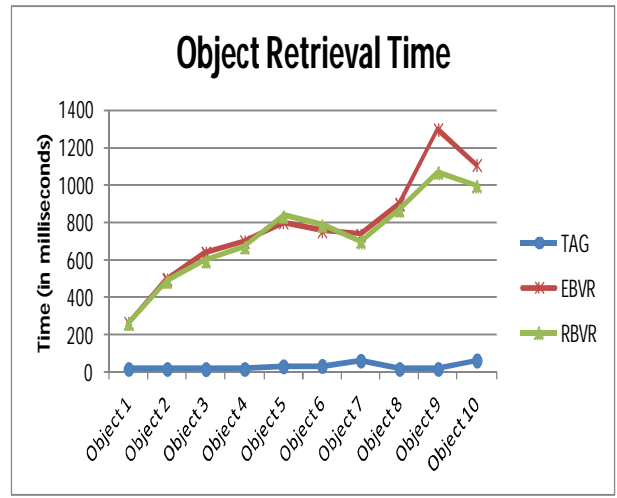

Fig. 4 Time taken to retrieve the object

### B. Single Version Retrieval

The performance is measured for retrieving a single earlier version of a document as the versions of the document gets increased (Fig. 5.). When a version is queried, EBVR always refers to its previous version, which in turn might refer to its previous version. In this way it takes more time to retrieve the queried version.

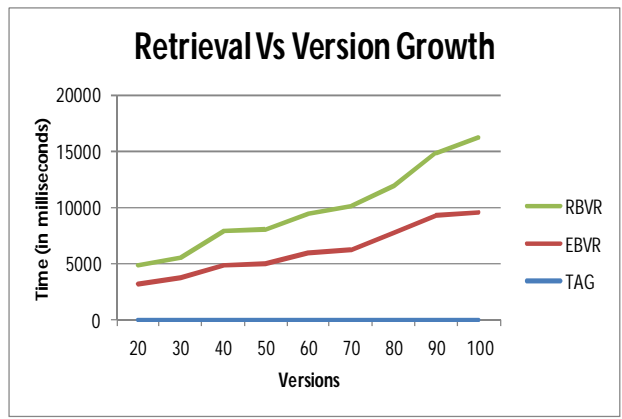

Fig. 5. Retrieval time against the growth of versions

The RBVR focuses on the common unchanged parts. Thus it takes less time when compared to EBVR. But this model also need additional efforts to construct the queried version like referring the Reference Record. In VTAG with the help of hash table, it is very easy to locate the queried version. Thus it is clear VTAG performs well compared to other models.

We randomly chosen 10 different versions of various documents. The size of the collection is fixed to 200 documents with 100 versions each. Fig. 6. shows the results when we studied the performance while the earlier version is retrieved.

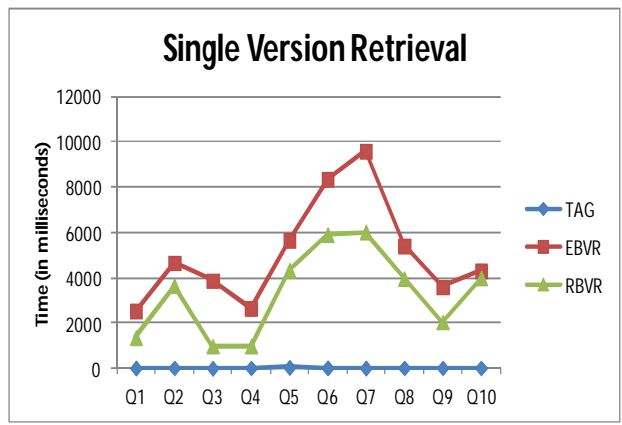

Fig. 6. Time elapsed to retrieve a single version

EBVR stores all the changes of either insertion or deletion. When an updation takes place, it records the deletion as well as the addition of the new information added. It concentrate on both changed and unchanged information. Thus the time taken to read the edit-script before retrieving (constructing) the queried version, is quite large. In VTAG, it needs additionally the time to access the hash table along with time to locate the document. Hence VTAG outperforms than the earlier model.

*C. Multi-version Retrieval*

This study focuses on retrieving more than one version of the document. Eight different versions of the various documents are queried. Figure. 7. shows the results of this experiment. Both EBVR and RBVR has to bring almost the maximum previous versions, required to construct the queried version, to main memory. Then it needs to process all these versions. VTAG directly accesses the version queried. Thus it consumes less amount of time and memory.

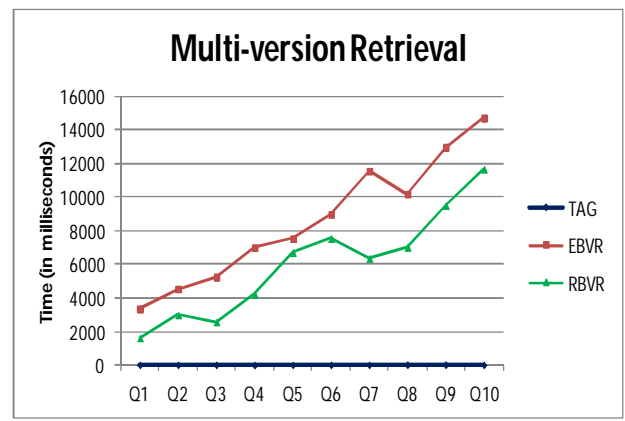

Fig. 7. Multiple versions of a single object

The main goal of these experiments is to study the retrieval performance of the VTAG model. The result of the study shows that VTAG model performance is better than the EBVR and RBVR model. The other operations such as insertion and deletion have also been analyzed and found that VTAG model outperforms the earlier models.

V. CONCLUSION

In this study a modified architecture for indexing the multi-versioned documents is proposed. From the results of the experiments it is shown that the proposed indexing technique performs well when compared to the other existing techniques. This technique can be applied to various multi-versioned publication systems in future to check the behavioural changes of the Indexer.

ACKNOWLEDGMENT

This study is a part of ICBR - Major Research Project (41- 642/2012(SR)) funded by University Grants Commission, India.

## REFERENCES

[1] R. M. Aliguliyev, "A new sentence similarity measure and sentence based extractive technique for automatic text summarization," *Expert Syst. Applic.,* vol. 36, pp. 7764-7772, May 2009.

[2] S. Chakrabarti, *Mining the Web: Discovering knowledge from hypertext data*. San Francisco: Morgan-Kaufmann, 2003.

[3] S. Y. Chien, V. Tsotras and C. Aniolo, "Efficient schemes for managing multiversion XML documents," *The VLDB Journal*, vol. 11, pp. 332–353, December 2002.

[4] Y. L. Chen and Y.T. Chiu, "An IPC-based vector space model for patent retrieval, " *Inform. Process. Manage.*, vol. 47, pp. 309-322, May 2011.

[5] R. L. Cilibrasi and P.M.B. Vitanyi, "The google similarity distance," *IEEE Trans. Knowl. Data Eng.,* vol. 19, pp. 370-383, March 2007.

[6] (2013) The Google Scholar. [Online]. Available: http://scholar.google.co.in/intl/en/scholar/about.html

[7] (2013) The IEEE Xplore website. [Online]. Available: http://ieeexplore.ieee.org/Xplorehelp/Help_start.html

[8] (2002) National Archives of Australia's Policy Statement Archiving Web Resources - A policy for keeping records of web-based activity in the commonwealth government. [Online]. Available: http://www.naa.gov.au/recordkeeping

[9] N. Ratprasartporn, J. Po, A. Cakmak, S. Bani-Ahmad and G. Ozsoyoglu, "Context-based literature digital collection search," *VLDB J.,* vol. 18, pp. 277-301, January 2009.

[10] M. Thangaraj and V. Gayathri, "A context-based technique using tag-tree for an effective retrieval from a digital literature collection," *Journal of Computer Science*, vol. 9, pp. 1602-1617, November 2013.

[11] (2002) WWW Distributed Authoring and Versioning (webdav) [Online]. Available: http://www.ietf.org/html.charters/webdav-charter.html